\documentclass[journal]{IEEEtran}

\usepackage{makecell}
\usepackage{longtable}
\usepackage{booktabs}
\usepackage{tabularray}
\usepackage{booktabs}
\usepackage{graphicx}
\usepackage[numbers]{natbib}

\AtBeginDocument{%
  }

\begin{document}

\title{A Comprehensive Review of Computational Models for Longitudinal Wearable Sensor Data Analysis}

\author{Paula Lago \\ \small{paula.lago@concordia.ca} \\ \small{Department of Electrical and Computer Engineering, School of Health} \\\small{Concordia University, Montreal, QC, Canada}}

\maketitle
\begin{abstract}
Wearable devices are increasingly used as tools for biomedical research, as the continuous stream of behavioral and physiological data they collect can provide insights about our health in everyday contexts. Long-term tracking, defined in the timescale of months of year, can provide insights of patterns and changes as indicators of health changes. These insights can make medicine and healthcare more predictive, preventive, personalized, and participative (The 4P's). However, the challenges in modeling, understanding and processing longitudinal data are a significant barrier to their adoption in research studies and clinical settings. 
In this paper, we review and discuss three models used to make sense of longitudinal data: routines, rhythms and stability metrics. We present the challenges associated with the processing and analysis of longitudinal wearable sensor data, with a special focus on how to handle the different temporal dynamics at various granularities. 
 We then discuss current limitations and identify directions for future work. This review  is essential to the advancement of computational modeling and analysis of longitudinal sensor data for pervasive healthcare.
\end{abstract}
\section{Introduction}
\label{sec:introduction}

Pervasive healthcare is transforming the way healthcare is delivered and managed, leveraging advancements in sensor technology and data analytics to provide continuous and personalized health monitoring in real-world environments~\cite{Smuck2021, Witt2019}. This shift towards pervasive healthcare necessitates a move beyond traditional, episodic healthcare models to a paradigm where continuous data streams from wearable and ambient sensors are used to gain insights into an individual's health and well-being~\cite{bardram2008pervasive}. The analysis of longitudinal data, spanning weeks, months, or even years, offers a unique and valuable perspective for realizing the full potential of pervasive healthcare. 

\begin{figure*}[htb]
    \centering
\includegraphics[width=\linewidth]{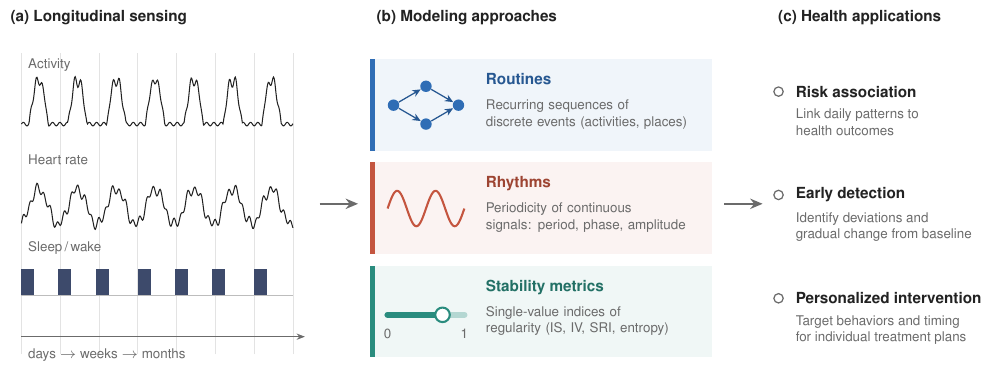}
    \caption{Overview of the review. (a)~Wearable and ambient sensors produce behavioral and physiological streams over days to months. (b)~Three families of models summarize these streams: routines (recurring sequences of discrete events), rhythms (periodicity of continuous signals), and stability metrics (single-value indices of regularity). (c)~The resulting patterns support risk association, early detection of change, and personalized interventions.}
    \label{fig:summary}
\end{figure*}

Unlike approaches that primarily focus on recognizing short-term patterns in sensor data, such as those commonly used for activity recognition~\cite{har} or stress detection~\cite{8115233}, longitudinal data analysis enables the identification of long-term trends and subtle changes in an individual's health and behavior~\cite{longtermselftracking}. These long-term insights are essential for understanding the progression of chronic conditions~\cite{sricardiovascular,obesityregularity}, evaluating the effectiveness of interventions over time~\cite{brainintervention2022}, and proactively identifying potential health risks before they escalate into acute events. For instance, analyzing variations in sleep patterns over several months could provide early indications of developing sleep disorders or mental health conditions~\cite{sung2024emerging}. Similarly, long-term monitoring of physical activity patterns could reveal a gradual decline in mobility, suggesting the need for interventions to prevent falls or functional decline~\cite{Cook2015}. Short-term pattern recognition, while valuable for real-time applications, may not capture these gradual, yet crucial, health changes that unfold over extended periods. By understanding an individual's baseline patterns and how they deviate over time, longitudinal data analysis can inform the development of personalized interventions.

A key challenge to fully leveraging the potential of longitudinal data analysis in pervasive healthcare lies in the inherent complexities of modeling and processing such data. Longitudinal data, collected over extended periods, often exhibits high dimensionality, capturing numerous variables from multiple sensors. This data abundance, while offering rich insights, necessitates sophisticated computational techniques to effectively extract meaningful information and manage the vast volume of data~\cite{Xu2021researchpractices}. Moreover, data quality can vary significantly due to sensor limitations, missing data points, and the influence of contextual factors, impacting the accuracy and reliability of analysis~\cite{Cho2021, bottcher2022data}. Supervised learning methods, while widely used in healthcare, encounter limitations when applied to longitudinal data due to data scarcity, quality issues, and heterogeneity~\cite{beyondsupervisedlearning}. Additionally, the temporal granularity of longitudinal data can vary considerably, requiring models that can handle different timescales and align events across diverse sensor streams. The lack of standards to analyze sensor data has made results obtained so far non-generalizable and difficult to replicate~\cite{Stupple2019}. These challenges, if unaddressed, can hinder the wider adoption of longitudinal data analysis in research and clinical settings, limiting the realization of pervasive healthcare's full potential.

This review focuses on three computational models used to analyze longitudinal sensor data: routines, rhythms, and stability metrics (Fig.~\ref{fig:summary} and Table~\ref{tab:summary}). Routines(Section~\ref{sec:routines}) are recurring patterns of discrete events, like daily living activities or locations visited. Routines help to understand the behavior patterns associated with health outcomes. Computational models of routines range from representing routines as probable sequences of events, capturing dependencies between events in the order they happen, to viewing them as sets of frequently co-occurring events within a specific time frame. Rhythms, in contrast, center on the periodicity and temporal regularity of physiological and behavioral processes~(Section~\ref{sec:rhythms}). A prime example is the circadian rhythm, governed by the internal biological clock and influenced by the day-night cycle. Rhythms can be modeled as sinusoidal functions (like cosine waves) to capture characteristics such as period, amplitude, and phase. Stability metrics(Section~\ref{sec:metrics})  quantify the strength of repetitiveness in behavioral or physiological patterns without explicitly modeling the patterns themselves. These single-value metrics offer a simplified way to assess the regularity of behavior. For instance, high stability in activity patterns might suggest a consistent daily routine, while low stability could indicate greater variability in day-to-day behavior.

While several reviews explore the use of smartphones and wearable sensors in healthcare, most focus on application areas, evidence for disease management, or technological advancements in passive sensing~\cite{Lu2020}. Fewer address the specific modeling approaches crucial for analyzing longitudinal patterns. For instance, Mohr et al. concentrate on mental health applications, providing a model for personal sensing and reviewing sensor-based detection of mental health conditions and behavioral markers~\cite{Mohr2017}. Similarly, Witt et al. detail algorithmic approaches for generating fitness and health indicators from wearable data, primarily focusing on present snapshots of user status rather than long-term trends~\cite{Witt2019}. While they briefly discuss behavior change detection techniques, these methods only touch upon longitudinal data analysis. This paper aims to fill this gap by providing a comprehensive review of models specifically designed for understanding longitudinal patterns in wearable sensor data, focusing on routines, rhythms, and stability metrics.

This review makes the following contributions: 
\begin{itemize}
    \item We provide a comprehensive review of existing methods for modeling behavior and predicting health outcomes using continuous, longitudinal sensor data. We focus on three key modeling approaches: routines (sequences of events), rhythms (periodic patterns), and stability metrics (quantifying pattern regularity), discussing the strengths and weaknesses of each method, to unify clinical and computational perspectives. 
    \item We provide an overview of common metrics used to quantify the regularity of patterns and classify these metrics based on whether they use an unimodal or multimodal approach. 
    \item We discuss open research challenges in the field, including choosing a model, determining window lengths, establishing baselines and analyzing change, model evaluation, data heterogeneity and quality, and the lack of context in analysis.
    \item We highlight the need for standards in longitudinal data analysis to improve the comparability of results and data sharing. We need standards for feature and event window length, data formatting, and missing data management to facilitate collaboration and generalization of findings.
\end{itemize}

This review explores routines, rhythms, and stability metrics as models to understand longitudinal health data. These models aim to capture the underlying temporal patterns in behavior and physiology over extended periods. We provide a comprehensive overview of the models, discussing their strengths, limitations, and open research challenges. By synthesizing the current state of the art in longitudinal data analysis, this review aims to guide future research and contribute to the development of more effective and clinically relevant applications of wearable sensor data in pervasive healthcare.

\section{Rhythms}
\label{sec:rhythms}
A rhythm is a cyclic pattern, repeated at regular temporal intervals. Human behavior and physiology follow biological rhythms, the most common being linked to the day/night cycle. These rhythms are governed by an internal pacemaker called the suprachiasmatic nucleus (SCN), an internal structure located in the brain, and are affected by external factors like light, temperature, and social factors, among others~\cite{albrecht2012timing,Reid2019}. Research has found strong links between the strength and alignment of this central clock with environmental and behavioral cycles and our physical and mental well-being~\cite{doherty2018circadian,lee2024real}. As wearables measure physiological outputs influenced by the internal circadian clock, they offer a promising opportunity to estimate the circadian phase in real-world settings. 
Characterizing the various behavioral and physiological rhythms from sensor data is considered an important next step in understanding these rhythms in everyday life conditions to design personalized diagnostics and interventions~\cite{abdullah2017circadian}.  

Longitudinal measures of activity and physiology obtained with passive sensing have been used to model the SCN rhythm or directly model the rhythms of activity, heart rate, or body temperature. Although the latter may not strongly correlate with the SCN rhythm, as they are affected by various other factors, their characterization can still provide information about health status~\cite{Bowman2021}. Moreover, the study of multiple rhythms can shed light on how these rhythms and their alignments affect our health. We now explain four rhythm models. 

\subsection{Template curves}
One simple way to model the typical rhythms of the data is to model a "template" curve for the day.  A template curve is created by averaging (or calculating the median value) across all days for each time slot in the day (e.g., every minute, hour). Activity curves~\cite{Dawadi2015} were proposed to model the probability that each activity occurs in each time interval when the measurements are not continuous signals but events. 

These curves can be used for anomaly detection if values deviate more than a certain amount from the template. The main issue is defining the time over which to calculate the template. They have also been used to obtain deviation from the template as a feature for classification purposes. 

While calculating template curves is simple, one of the major drawbacks of curves is the lack of context. A single template curve cannot account for differences due to day-to-day variations or monthly variations. Of course, template curves could be created for each day of the week or every day of the year, as is done with weather data, but they could lose representational power. 

\subsection{Cosine models}
Rhythms are most commonly modeled as sinusoidal functions, characterized by multiple parameters. The first parameter is its period. As such, circadian rhythms are those with a period of about 24~h (\emph{circa diem}; the average intrinsic circadian period in humans is about 24.2~h), while an ultradian (infradian) rhythm is one with a period substantially shorter (longer) than a day~\cite{volpato2005biological}. While 24 hours is usually assumed for the period, due to its relevance for the internal master clock, frequency analysis methods like the Fourier Transform, wavelets, spectral analysis, and periodograms can be used to extract personalized periods from the longitudinal sensor data. This period analysis can help uncover individual variations, as well as various rhythms. 

\begin{figure}
    \centering
    \includegraphics[width=\linewidth]{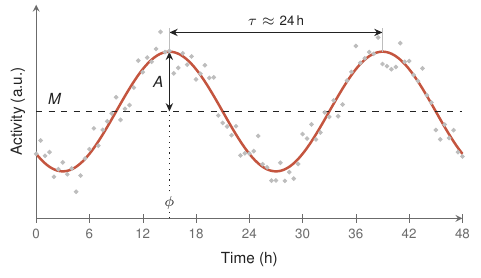}
    \caption{Single-component cosinor model fitted to illustrative hourly activity data (gray). The fitted curve (red) is characterized by its MESOR ($M$, rhythm-adjusted mean), amplitude ($A$, half the peak-to-trough difference), acrophase ($\phi$, time of the peak), and period ($\tau$, approximately 24\,h for circadian rhythms).}
    \label{fig:cosinor}
\end{figure}

By modeling rhythms as a sinusoid, the amplitude, MESOR, and Phase describe the rhythm of behavioral or physiological signals (Figure~\ref{fig:cosinor}). Phase measurements, given in hours, are of particular relevance as they enable the description of the relative timing of physiological rhythms with environmental and behavioral cycles. These metrics are obtained by fitting the sensor data to a cosine wave of a predefined period using the least square method --- this is called the COSINOR method~\cite{cosinorrhythmometry}. Equation~\ref{eq:cosinor} shows a single component COSINOR, where M is the mesor, A is the amplitude, $\phi$ is the phase, $\tau$ is the period, and $e(t)$ is the error term. 
 Other models include additional components to consider, for instance, the effect of light in the rhythm. Interested readers are referred to~\cite{Forger2017} for a complete discussion of circadian rhythm models.   

\begin{equation}
    Y(t) = M + Acos(\frac{2\pi*t}{\tau} + \phi ) + e(t) 
    \label{eq:cosinor}
\end{equation}

The model is used to characterize rhythms of single signals, for instance, activity or heart rate. In these cases, the model focus more on how these processes respond to systemic cues and their relationship with the SCN.

While characterizing the rhythms of physiological and behavioral processes is important, the importance of understanding the relation between multiple rhythms should not be disregarded. Yan et al.~\cite{Afsaneh2022} proposed a framework that combines period finding with COSINOR to independently describe rhythms of various signals and then uses these descriptors to predict depression risk. Liu et al.~\cite{Runze2024} also models a cosine rhythm for each signal and finds that rhythmic models are correlated with productivity. While these are  preliminary results in a specific population, they show the potential of combining multiple rhythms in the development of digital health services.  

\subsection{Models to predict SCN rhythm}
As previously mentioned, one common goal of modeling rhythms through sensor data is to model the SCN rhythm: predict its period, phase, and amplitude. The gold standard for measuring the SCN rhythm is taking multiple samples of melatonin, cortisol, or core body temperature under highly controlled conditions such as dim light or a constant routine protocol~\cite{Dijk2020}. These procedures are costly and invasive, so passive sensing provides a cost-effective and ecological way to predict it. 

The COSINOR method has been used with actigraphy (accelerometer-based) and physiological data to model the SCN rhythm with relative accuracy in ambulatory settings~\cite{cornelissen2014cosinor}. To account for influences on the measurement other than the circadian rhythm, various mathematical models of the clock have been proposed which model the oscillatory nature of the clock and how external light input shifts its phase~\cite{Forger2017, kim2023wearable}. Wearable data is used to estimate light exposure.

In most cases, though, passive sensor data is used to predict dim-light melatonin onset (DLMO), the time at which melatonin secretion begins to rise in the evening under dim-light conditions, which is a gold-standard marker of SCN phase~\cite{Dijk2020}. Deep learning and traditional machine learning techniques involving sensor measurements of activity, skin temperature, heart rate, and other physiological processes have been proposed for this purpose~\cite{Wan2021, Brown2021}. While these methods are promising, other studies have shown that the error could be higher in populations who may benefit the most from precise metrics such as shift workers, jet-lagged people, and people with sleep disorders~\cite{Wu2022}. Novel methods still need to be evaluated with more diverse and specific populations.  

Characterizing the SCN rhythm is pivotal to further our understanding of its role in health outcomes, however, researchers also acknowledge that different processes may exhibit different rhythms which need not have the same characteristics as other markers such as the melatonin onset (DLMO). Their characterization is useful for understanding underlying processes and might further increase our understanding of the role of the alignment of various rhythms with health. The methods to analyze such rhythms are similar to those described above, but models may consider the effects of confounder factors. For instance, to analyze the circadian rhythm of heart rate, Bowman et al.~\cite{Bowman2021} first removes the effects of activity, to uncover the underlying rhythm.

\subsection{Rhythm Models excluding Cosine Regression}
Since sensor data may not always follow a cosine waveform, non-parametric metrics have also been used to describe the rhythmic properties of sensor data, particularly activity data. These metrics are explained in Section~\ref{sec:metrics}. Non-negative matrix factorization (NMF) has also been proposed to discover typical daily rhythms~\cite{Aledavood2022} and multiday~\cite{Ceolini2023} rhythms without the need for a regression. A dataset of the daily or multi-day sensor signals is created where each column represents an individual's data, which can be daily, weekly, or monthly data. NMF factorizes this dataset (a matrix) to find a small number of common patterns, such that each individual's data can be represented as a linear combination of the patterns. This method is also used for routine modeling (Section~\ref{sec:routines}) by creating a matrix of events.

\section{Routines}
\label{sec:routines}
One approach to modeling long-term patterns is through routines. A routine is generally thought of as a sequence or set of actions performed regularly. 
While there is no agreed formal definition of a routine, the term is usually used to describe recurring patterns of discrete \textit{events} such as daily living activities, work tasks, or visited locations. As such, two main dimensions encompass a routine: co-occurring or sequences of events and their temporal recurrence. In contrast to rhythms that focus on the periodicity of sensor measurements, routines focus on the frequent repetition of events.   

Interest in computationally modeling routines is longstanding. Their computational study first emerged as a way to support everyday tasks: routines enable the prediction of the next event, which helps a system provide effective recommendations, assistance, or information for that event. Understanding routines from sensor data has been studied to analyze daily life routines~\cite{Ranvier2015}, family routines~\cite{Davidoff2010,Banovic2016}, work-place routines~\cite{Brdiczka2011,Su2013,Swain2019}, and travel routines~\cite{Liao2007,travel2020}. In the context of healthcare applications, routines help understand behavioral patterns that can be associated with health outcomes either positively or negatively~\cite{flexibility2022}.

\subsection{Models of routine}
\begin{figure*}
    \centering
    \includegraphics[width=\linewidth]{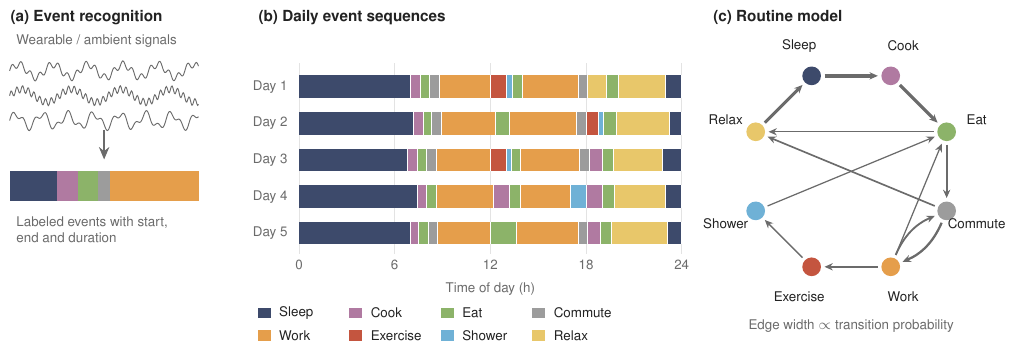}
    \caption{Routines are often modeled as networks of recurring events. (a)~Application-relevant events are recognized from wearable or ambient sensor signals. (b)~Each day becomes a sequence of events with start times and durations. (c)~Aggregating sequences over many days yields a routine model in which edge width represents the probability of the transition between consecutive events.}
    \label{fig:routines}
\end{figure*}

Routines are usually modeled as repetitive sequences of events (Figure~\ref{fig:routines}). As such, to discover routines from sensor data, it is first required to define and identify events. What makes a sequence "repetitive" is considered from multiple points of view: the frequency of occurrence (frequentist approach), the likelihood of occurrence (bayesian approach), and the periodicity of occurrence.  

\subsubsection{Defining events}
In the context of identifying routines from sensor data, an event can be defined as a significant occurrence or activity that can be detected or inferred from the data collected by sensors. It represents a specific action, behavior, or state of the individual being monitored.  An important characteristic of events is their duration, that is, they are not instantaneous points in time, but rather an entity with a start and end time. In general, an event characterizes what is being done, by whom, when, and where. 

Events can vary widely depending on the context and the type of sensor data being analyzed. For example, in the context of daily activities, events could include actions like cooking, walking, driving, sleeping, or working out~\cite{behaviorome2021} while in the context of travel routines, events correspond to locations visited~\cite{travel2020}. Commonly, an event is described by a set of sensor readings obtained during a time window which may include changes in sensor readings, such as motion, location, temperature, or other relevant parameters, which correspond to the occurrence of the event and can be used to classify them~\cite{Banovic2016}. 

Defining and identifying events can be a difficult task. Event definition depends on the sensor technology used and the probable duration of events. In some cases, events are defined directly by the sensor readings obtained during a window of time. In such cases, the number of events can be explosive, and unsupervised learning techniques are used to group them into a more manageable set (see next subsection). Events can also be identified by a \texttt{label}, requiring supervised learning to identify them. The activity recognition literature deals with this problem and has been extensively reviewed elsewhere~\cite{Lara2013, Chen2012, Wang2019, Chen2022, Dentamaro2024}.

\subsubsection{Routines as highly probable events}
Initial concepts of routine were based on daily routine, that is, how the day unfolds. The behavior of a particular day can be expressed as a vector $b$, describing the event that occurred at each particular time. This vector can be categorical, that is, each item bt identifies the event that occurred at time t as a category; or binary, that is, the item $b_{jt}$ has the value of 1 if the event $j$ occurred at time $t$. A matrix of multiple days $\mathcal{B} = D\times b$   describes the behavior over some time. 

From this matrix, routines can be extracted. The concept of eigen-behaviors, representing the eigenvectors of this matrix, was used by Eagle et al.~\cite{Eagle2009} to find common behaviors in a day. An eigen-behavior represents an intrinsic behavior pattern, eigenbehaviors, then, can be used to describe the overall routine of an individual as any particular day can be described as a linear combination of eigenbehaviors. While a predefined 24-hour period was assumed in this work, different behavior vectors were found for different situations showing that ``eigenbehaviors'' describe both typical patterns and variations of them.  

Similar to this approach, topic models, a technique of natural language processing to automatically discover topics in a document collection, were used to find "topics" of behaviors~\cite{Huynh2008, castanedo2014learning, Rieping2014}. In these models, an event is considered a word, each day is a document and a set of days is a document collection. A topic describes a group of behaviors that typically occur together in a day, and each day can be described as a combination of topics. 

One drawback of the previous methods is that they are based on individual data, suffering potentially from a cold-start problem. Only when there is sufficient data from a person, can we learn their routines and make inferences from their behavior patterns. To overcome this drawback, a collaborative-filtering-based approach was proposed by Xu. et al\cite{Xu2021depression}, which combines population data with personal data to find relevant behavioral features to infer depression.  

The idea behind these techniques is that there is a set of behaviors occurring at a particular time \textit{within predefined duration bins} that occur repeatedly. While a particular day exhibits a different combination of these typical patterns, the emerging patterns describe the routines of a person and can be used to represent any particular day. 

The problem with this view is the lack of sequence representation. Each day is described as a ``bag-of-events'' with positional information. While this representation captures some of the sequential context, it cannot capture partial orders, such as events happening next to each other regardless of the time. For instance, ``eigenbehaviors'' can find patterns such as a late wake-up leading to a no-work day, it is harder to find sequences of activities that can happen together at no particular time. For example, a routine might consist of gym and then eating, but it can occur at different times of the day. Or if there is a sequence of office, gym, and then going home, always starting at 5 pm but having different durations for each depending on traffic, day, and others, it might be hard to discover due to the dependence on time and the predefined time bins. 

\subsubsection{Routines as highly probable sequences}
Sequential models analyze data in the order of arrival, thus, capturing dependencies among the events. Both probabilistic and frequentist approaches have been used to model routines as sequences.  From a frequentist perspective, routines are events that empirically occur frequently within a given timeframe based on repeated observations, while from a Bayesian perspective, routines involve incorporating prior beliefs or knowledge about the likelihood of events occurring and updating these beliefs with new evidence.

Frequent-sequential pattern mining and association rule mining~\cite{Aztiria2013, Ranvier2015, Xu2019, alia2024daily} algorithms work under the frequentist assumption to identify events occurring repeatedly together in an event log. The extracted routine is then expressed as rules~\cite{Xu2019}, action maps~\cite{Aztiria2013}, automatas~\cite{li2018automatic}, or as a grammar~\cite{Li2015}.
Frequent sequences can represent typical sequences of events in the routine of a person but it is harder to model the uncertainty of multiple possible sequences. Another approach was proposed to consider context-based sequences~\cite{LAGO2019191}, to model frequent sequences that depend on the context such as day of the week or the weather, independently of the frequency of that context. 

From a probabilistic perspective, Bayesian Networks, which include hidden Markov models, have been used to learn and model the sequential dependencies of events~\cite{Liao2007,  Enshaeifar2018, Chifu2022}. These networks model the transition probabilities between events, as well as the correlation between events and sensor readings. A common problem is defining the structure of the network to learn the parameters. Banovic et al.~\cite{Banovic2016} proposed a model based on reinforcement learning to learn the structure. The learned network of events describes the behavior of the person. By considering the context, such as location or time, in the description of the event, variations in behavior are modeled as well. However, the combinatorial explosion of the number of events and the need to manually describe them make it a computationally expensive model.

Both models have been used extensively in research, their main strength being the possibility of inspecting the routine and knowing exactly what is the next event, making them appropriate for intervention design. Their other strength is the possibility of modeling contextual changes. This is done either by adding it to the event definition or as part of the mining processing. In sequential patterns, modeling includes deciding the maximum length of the sequence and whether gaps or different orderings can be included. These decisions may affect the final learned routine. 

Deep learning models have also been used to model longitudinal sensor data. For this, the events are defined by slices of data which can be a day of sensor events. Due to their black-box nature, Deep learning models have mostly been used in tasks to associate behavior with health outcomes. Recently, Xu et al.~\cite{Xu2023} proposed a 1d Convolutional network to detect depression, which also learns the continuity of behavior. This can help learn the behavioral patterns but with the drawback of not being interpretable.
While deep neural networks such as LSTM and GRU have been used to model other sequential data, their use to model routine from sensor data has been limited.  

As mentioned before, models of routines are based on the definition and identification of events. This includes multiple assumptions about the duration of events, often left as predefined bins or windows, which may not accurately represent the duration of all activities leading to errors, and about the accuracy of identification. The propagation of errors from the identification of events to the modeling of routines can affect the usefulness of the model. While the events have also been defined as sensor event counts or statistics over predefined windows of time, this can lead to a combinatorial explosion in the number of possible states or require manual definition of thresholds to limit the possibilities. This is still a major drawback for routine analysis.

\section{Quantifying behavioral stability with longitudinal sensor data}
\label{sec:metrics}

Both routines and rhythms model long-term sensor data, focusing on different characteristics of the patterns (periodicity, frequency, sequential dependencies). Other approaches focus on quantifying the \textit{strength} of repetitiveness without modeling the specific patterns.
A person with high strength (or stability) can be thought of as having a more regular or strict routine while a person with lower strength will exhibit more variability in their day-to-day life. While these metrics can't describe the patterns, they characterize the repetitiveness of behavior. 
The simplicity of a single metric makes metrics attractive to perform statistical analysis, as a tool for early screening, and because they can be more easily interpretable than other models. For instance, less stable and more fragmented rhythms have been associated with depression, higher BMI, and smoking~\cite{Luik2013-lv}. 
Research focuses on finding the relation between metrics and physical, mental, or cognitive health.  These metrics can be classified as being uni-modal if they consider a single sensor modality, or multi-modal if they consider multiple sensor modalities at the same time. A summary of metrics is presented in Table~\ref{tab:metric}, and they are discussed below.

\subsection{Rhythm-based metrics}
Rhythm-based metrics use a single sensor modality to obtain a measure of the strength of the rhythmicity of behavior, that is, how similar the measurements taken at the same over multiple days are. The interday stability, intraday variability,  and relative amplitude (RA) metrics come from circadian rhythm research, as non-parametric measures of the strength of the circadian rhythm and are most commonly calculated using activity counts from actigraphy devices. They measure the strength, fragmentation, and amplitude of the rhythm, respectively. 

The regularity and stability indexes have been proposed to measure the regularity from a variety of sensor modalities, but still consider only one modality at a time. They focus on measuring the similarity of events happening at the same time over multiple days. 
The regularity index measures how similar measurements taken at the same hour over two different days are, while the stability index compares the normalized cumulative distribution value at each hour over two days and then offers a value for a period by comparing all days to each other. 

These metrics rely on having an aggregated measure, usually hourly, to compare the value across multiple days. While the aggregation could be in smaller segments, it is not clear how the aggregation might affect the result. 

\subsection{Event based metrics}
Instead of relying on the measurements directly, event-based metrics rely on the detection of an event ---i.e., an activity, a location--- and then calculating the regularity of occurrences of this event over some time. For instance, the standard deviation in start/end time or duration of the event is a measure of how variable that event behavior is. The sleep regularity index focuses exclusively on the regularity of sleep schedule by measuring if the individual is asleep at the same times over the period being examined. Entropy measures the extent of repetition of events, regardless of their order, considering X as a random variable representing the activities in a day. For a person with a more stable routine, this random variable will have lower entropy, as the events are generated by a more regular process. Entropy measure is chosen for event-based analysis as the events are a categorical variable. The flexible regularity index~\cite{Wang2018} measures the edit distance between two days. By representing each event with a letter and considering the daily sequence of events as a character sequence, the edit distance counts the minimum number of insertions, deletions and substitutions needed to turn one day's sequence into the other's; smaller distances indicate more similar, and therefore more regular, days.

\clearpage
\onecolumn
\begin{longtable}{p{2cm}p{5.5cm}c} 
\toprule
         \textbf{Metric} & \textbf{Description} & \textbf{Formula} \\[6pt] \toprule \endhead
         \multicolumn{3}{c}{\textbf{\textit{Rhythm-based metrics}}}\\[6pt]
         \midrule
         Interdaily Stability (IS)~\cite{fischer2021} & A measure of how stable a person’s rhythm is over multiple days, calculated from $n$ samples of activity levels $x$, aggregated hourly $q$ hour-bins. $\bar{x_h}$ represents the average activity during hour $h$. IS scores range from 0, indicative of a total lack of rhythm, to 1, indicative of perfectly stable rhythm & $IS= \frac{n \sum_{h=1}^q (\bar{x_h} - \bar{x})^2}{q \sum_{i=1}^n(x_i-\bar{x})^2}$\\
         Intradaily Variability (IV)~\cite{fischer2021} & A measure of the fragmentation of activity or how activity level shifts between two consecutive time slots calculated from $n$ samples of activity levels $x$, aggregated hourly $q$ hour-bins. IV scores range of 0 to 2 with higher values indicating higher fragmentation & $IV= \frac{n \sum_{i=2}^n (x_i - x_{i-1})^2}{(n-1) \sum_{i=1}^n(x_i-\bar{x})^2}$\\
         Rest-activity Amplitude~\cite{WITTING1990563} & Measures the difference between a person’s daily activity (measured as the mean activity level over the most active 10 consecutive hours (M10)) and nocturnal activity (measured as the mean activity during the least active 5 consecutive hours (L5)) &$RA = \frac{M10 - L5}{M10 + L5}$ \\
         Regularity Index~\cite{behaviorome2021, Wang2018,brainintervention2022} & Compares the uniformity of a person’s schedule, comparing hourly (t) data from two days (a,b). Sensor measurements ($x_a^t$) are normalized to (-0.5, 0.5) & $RI_{a,b} = \sum^T_{t=1} \frac{x_t^a x_t^b}{T}$\\
         Stability Index~\cite{schizophrenia2020} & Measures the inverse of the median across a period of multiple days P  of the distances between the normalized cumulative activity distributions ($C_b^{d_1}$) on two different days. It characterizes the degree to which behavioral patterns varied across the period. A higher index means higher stability.  & \makecell{\\$SI_b (P)= 1 - median ( D_b(i,j )|i,j \in P , i \neq j $ \\ where $D_b(d_1,d_2) =  \frac{1}{M}\sum_{i=1}^M |C_b^{d_1} - C_b^{d_2}|$}\\
         Circadian Rhythm ~\cite{Wang2018}&Measures the strength with which a user follows a 24-hour rhythm in behaviors by calculating the ratio of the energy that fall into the $24 \pm 0.5$ h period over the total spectrum energy in the $24 \pm 12$ h period. Here, psd(x) denotes the power spectral density at frequency bin x. & $ CR = \frac{\int_{2\pi/24.5}^{2\pi/23.5}psd(x)dx}{\int_{2\pi/36}^{2\pi/12}psd(x)dx} $\\
         \\[6pt]
         \midrule
         \multicolumn{3}{c}{\textbf{\textit{Routine-based metrics}}}\\[6pt]
         \midrule
         Standard deviation~\cite{depression2018,Wang2018} & Measures the within-individual variability in a derived parameter (x) of an event. For example, the start time or duration of the event. A higher value indicates more irregular behavior. & $s = \sqrt{\frac{1}{N-1} \sum_{i=1}^N (x_i - \overline{x})^2}$ \\
         Sleep regularity index~\cite{Phillips2017} & The probability of an individual being in the same state (asleep or awake) at any two time points 24 hours apart. $s_{i,j}$ is the state on day $i$ at epoch $j$, $M$ is the number of epochs per day, $N$ is the number of days, and $\delta=1$ if the two states are equal and 0 otherwise. SRI ranges from $-100$ to 100: 0 represents a random sleep schedule and 100 a perfectly periodic one. & \makecell{$SRI = -100 + \frac{200}{M (N - 1)}*$\\   $\sum_{j=1}^{M} \sum_{i=1}^{N-1} \delta (s_{i,j},s_{i+1,j})$} \\
         Flexible Regularity Index~\cite{Wang2018}& Edit-based distance to measure the difference between two days by counting the number of operations (insert, delete, substitute) to convert one day's events into the other day's events. Higher values indicate more dissimilar days. & See reference. \\
         Entropy~\cite{travel2020,Shi2024} & Measures the repetition of isolated events (i.e. travel events, activities), where E is the set of all possible events. A value of zero indicates no uncertainty (more stability); the maximum, $\log_2|E|$ (1 if normalised by $\log_2|E|$), indicates that all events are equally likely & $H(X) = -\sum_{x \in E} p(x)\log_2 p(x)$\\
          [6pt]
         \midrule
         \multicolumn{3}{c}{\textbf{\textit{Multimodal metrics}}}\\[6pt]
         \midrule
         Entropy Rate of a Sequence~\cite{travel2020}& Measures the extent to which ordered subsequences of events repeat over time. $p_n$ denotes the joint probability distribution of a subsequence of length $n$ &\makecell{ $H(X) =$ \\ $\lim_{n\to\infty} - \sum_{x_1^n \in E^n} p_n(x_1^n)\log_2\frac{p_n(x_1^n)}{p_{n-1}(x_1^{n-1})}$ }\\
         Multiscale Fuzzy Entropy~\cite{Reinertsen2017}& Measures the complexity of a time series& See reference. \\
         \bottomrule  

\caption{Behavior stability/variability metrics}
\label{tab:metric}
\end{longtable}
\clearpage
\twocolumn
 
\subsection{Multimodal metrics}
Previous metrics measure the stability of behavior considering one measured variable, but they can't measure relations across various behavioral and physiological measures such as sleep, steps, and heart rate. Amon et al.~\cite{flexibility2022} proposed an approach based on Multidimensional Recurrence Quantification Analysis (MdRQA) to provide a metric of the stability of behavior from multiple sensor signals. This method quantifies the interactions between multiple signals, thus offering a more holistic view of the repetition of patterns, which can also be understood as contextualizing the signals and their regularity. Similarly, the entropy rate~\cite{travel2020} looks at the regularity of multiple events instead of an isolated event by considering the sequence of events.

\subsection{Metrics summary}
Metrics offer a valuable way to quantify the regularity and stability of patterns, overcoming some of the limitations of routines and rhythms. By providing a single, quantifiable marker, metrics can simplify the analysis of complex longitudinal data and facilitate the identification of associations with health outcomes. For instance, metrics like the standard deviation of event timings or the sleep regularity index can be easily correlated with conditions like depression, schizophrenia, or cognitive decline, helping researchers understand the behavioral underpinnings of these conditions.

However, it is important to note that most metrics rely on a single signal modality and a specific timeframe, which can restrict the scope of analysis. While researchers commonly calculate metrics for different signal modalities and timeframes to gain a broader perspective, multimodal metrics that capture the interplay between various behavioral and physiological signals could offer a more holistic view. Furthermore, the meaning of stability can vary across different signals, with greater stability being beneficial for some outcomes while variability might be more advantageous for others~\cite{schizophrenia2020}. A deeper understanding of the relationships between stability, variability, and health outcomes across multiple signals is crucial for developing more nuanced and effective interventions.

\section{Discussion}
\label{sec:discussion}
Having examined three approaches to analyzing longitudinal sensor data, we now move on to discussing some open research challenges. 

\subsection{Choosing a model}
Table~\ref{tab:summary} presents a summary of the models we have reviewed. 
The first key open question is how to choose a model for long-term sensor data analysis. A key consideration is the goal of the patterns. Common uses include (1) to find associations with health outcomes, to which metrics have an appeal due to their simplicity; (2) to provide assistance and predictions of the next events to which routines may be more suited; and (3) to find change and deviations, to which possibly all three models may help. 

A combination of models is not uncommon, especially in conjunction with machine learning and deep learning for all three tasks. For this, the following open issues need to be solved as well.

\begin{table*}[t]
\centering
\caption{Comparison of Longitudinal Data Analysis Models}
\label{tab:summary}
\begin{tabular}{|c|l|l|l|}
\hline
\textbf{Feature} & \textbf{Routines} & \textbf{Rhythms} & \textbf{Regularity Metrics} \\
\hline
\textbf{Focus} & Sequences and co-occurrence of  & Periodicity and regularity of & Quantifying the strength of   \\
 & discrete events & continuous processes & repetitiveness in behavioral or  \\
 & & & physiological patterns without  \\
 & & & explicitly modeling the patterns \\
 & & & themselves \\ 
\hline
\textbf{Modeling} & Frequent sequences, probabilistic & Sinusoidal functions  & Statistical summaries or indices \\
 & models, deep learning & (e.g., cosine waves) & \\
\hline
\textbf{Applications} & Understanding behavioral  & Identifying disruptions in & Assessing the regularity of behavior  \\
 & patterns, designing interventions, & biological clocks, understanding &  for correlations with health outcomes, \\
 & anomaly detection & physiological processes, & early screening, and simplified \\
 & & health monitoring & interpretations \\
\hline
\textbf{Examples} & - Identifying a sequence of  & - Analyzing sleep-wake cycles & - \textbf{Interday Stability (IS):} Measures  \\
 & "wake up," "eat breakfast,"  & - Characterizing heart rate  & the stability of a rhythm over multiple days \\
 &  "go to work"  & variability throughout the day & - \textbf{Sleep Regularity Index (SRI):} \\
 & - Detecting deviations from & & Quantifies the consistency of sleep schedules \\
 & established routines & & - \textbf{Entropy:} Assesses the  \\
 & & & predictability of event sequences \\
\hline
\end{tabular}
\end{table*}

\subsection{Windows and window length}
During the analysis of longitudinal sensor data, an important consideration is the amount of time (window length) to analyze. It is important to recognize two different types of windows, neither more important than the other: one window to aggregate sensor data as features or events and another to analyze the patterns in sensor data. 

The first type of window, the event window, defines the length of time that is used to recognize an event or a feature (such as the number of steps taken or the average heart rate) that will later be used to obtain patterns in the form of routines, rhythm, or (ir)regularity metrics. This window is often defined in the range of seconds, a few minutes, or one hour. Extensive reviews have looked at the impact of the size of this window, for example, in activity recognition~\cite{Banos2014, Hiremath2021}. Automatic segmentation has also been studied for this problem, where the window size is automatically determined by changes in the data distribution~\cite{Deldari2020}. 

The second type of window, the pattern window, considers how historical data is used to analyze the patterns, for instance, 7 days or 1 month. Routines, rhythm features, and metrics might change depending on this length. Very little work has been done in the analysis of the impact of this length or how to define it. Most commonly, multiple window lengths are evaluated to find the one with the most predictive value for the problem at hand~\cite{Reinertsen2017, schizophrenia2020}. In general, shorter window lengths will have a narrow but current view of the person's patterns, while longer windows will have a wider view but it could be influenced by outdated patterns--- the weight of history is heavy. 

A lack of standards in the size of windows hinders the comparability of results and efforts for data sharing. It is not uncommon to find datasets that aggregate data in a too coarse way, making it harder for others to evaluate or compare results that were obtained with smaller windows. 

\subsection{Baseline establishment and change analysis}
Related to the problem of determining the pattern window length is the issue of change analysis. Ideally, when analyzing longitudinal data, one is interested not only in the patterns found in data but also in identifying if any pattern changes have occurred. It is, after all, changes that may be more correlated with health changes or health outcomes after an intervention~\cite{7565567, Chen2023}. 
However, most of the research so far focuses on analyzing the current behavioral patterns and the predictive power of these patterns for the health outcome being studied, and there is less research on how to analyze changes in behavior \textit{across} windows. 

Routine analysis has been used to identify changes in routine and anomalies by comparing current behavior with the learned models~\cite{LAGO2019191,Eisa2017}. To analyze changes in phase or amplitude of circadian rhythms over time, Bayesian spectral analysis and detrended fluctuation analysis have been used~\cite{mathmodelcircadian}. Changes are also detected statistically~\cite{Merilahti2016}. Some approaches based on \textit{distribution changes} evaluate whether behavior changed in two different windows of time by comparing the distribution of sensor data of one window to that of the other window~\cite{Sprint2018,brainintervention2022}. The work by Cook et al.~\cite{brainintervention2022} determines behavior change over two time windows as a probability distance. By defining the average probability that an activity occurs during a time interval (i.e. an hour of the day) over the time window, it defines behavior as a probability distribution for each time interval. It then calculates the distance between two time windows using the KL divergence of the two distributions.

To quantify the change, \cite{depression2018} uses the slope of the data to quantify how values are changing inside a given window, but it does not quantify if the stability/rhythmicity is changing over various windows. Cook et al.~\cite{brainintervention2022} use a  permutation-based test to determine the statistical significance of detected changes. Another approach to quantify change is to calculate the z-score of the measurement with respect to the baseline window and use a threshold of change to identify substantial change~\cite{mishra2020pre,lim2024accurately}. The z-score measures the deviation from the baseline, enabling comparisons across different days, individuals, and periods.

The problem of detecting and quantifying behavior or rhythm change can also be framed as a change point detection problem. While there are multiple time series change point detection methods~\cite{Aminikhanghahi2017-re}, these methods usually focus on the raw sensor values, which can be less meaningful for wearable sensor data, as the changes might be too granular or too fast. For instance, a changing heart rate due to exercise. For longitudinal data, the relevant changes span weeks or months of baseline pattern.

Other works frame the problem as an anomaly detection problem~\cite{Dahmen2021, Ye2015, Arifoglu2019, Riboni2016}. Anomaly detection algorithms analyze deviations from current patterns to detect abnormal activities, heart rate, or other abnormal values, but they do not focus on identifying if the baseline patterns are changing. In other words, while they detect a momentary deviation, they assume the overall pattern remains constant. 

Overall longitudinal pattern change analysis remains elusive, due to the difficulties in obtaining enough data for the analysis, the difficulties in defining time windows for the analysis, and the difficulties related to the evaluation of any such approaches.

\subsection{Model Evaluation}
Evaluating how accurately the routine, rhythm, or behavioral stability of a person is, is undoubtedly a difficult task. Unlike other tasks, such as supervised classification of activities, there is no ground truth data that can be used to compare the model. 
Routines have been evaluated by letting the user review the output and agree/disagree on it. However, this is still a subjective evaluation. For rhythms, comparison with rhythm parameters of the melatonin or cortisol provides an error measure, but as was discussed, this can be misleading, as different processes may have different rhythms.  

Recently, the evaluation focus has shifted to predictive power, that is, how well the patterns or metrics can detect or predict a health outcome (hospitalization, schizophrenia, depression, etc). These evaluations are important, as they focus on the clinical importance instead of computational metrics, but they are also elusive as they depend on the population in the sample, and have proven difficult to generalize. Moreover, they are difficult to translate to other uses of longitudinal pattern analysis, such as providing assistance with a task or identifying abnormal patterns. 

\subsection{Data Heterogeneity and Data Quality}
Related to the evaluation and generalization of results is the \textbf{data heterogeneity} issue. Different platforms, vendors, form factors, users, and software libraries result in significant differences in data~\cite{Stinsen2015}. These differences can result in very different pattern estimations, whether in routines or rhythms.  While metrics are developed and hand-crafted to find associations with health outcomes, their values can differ significantly due to the high heterogeneity, making it hard to generalize across different domains (sensing platforms, population groups, time)~\cite{VOS2023105026,Zhai2024-jw}. 

For this reason, deep learning methods have been proposed as a way to find more stable features in the sensor data to classify and predict health outcomes that could have behavioral symptoms mainly with multi-task learning to learn patterns in the different domains~\cite{10.1145/3659597,Peng2018,Samyoun2022,10.1145/3328932}. However, the generalization issue persists. 

A second issue for data quality is the inevitability of missing data~\cite{Thomas2022}. When data is being collected in natural settings, missing data may come due to battery issues, non-wearing events, or due to platform restrictions~\cite{Wang2018}. 

Proposals to deal with missing data include interpolation techniques and discarding data that contains missing data over a threshold percentage~\cite{Afsaneh2022}. There are no standards set for this and the correct management of missing data is still an open research issue. 

\subsection{Context}
Another issue in the current analysis methods is the lack of context in most of them. Several contextual factors can affect routine, rhythms, and the different metrics, including weekends, holidays, weather, or family visits. Not considering this can result in biased metrics, misunderstanding of the patterns, or a lack of insights into what causes changes in stability.  
Some models of routine consider contextual variations as a way to explain routine deviations and differentiate them from changes in behavioral patterns~\cite{elbert2011approach, LAGO2019191}. Some metrics, such as the regularity index, consider the day-to-day similarity, which may consider differences due to day-of-week changes, but not for other contextual factors. As was mentioned, this is also considered in the calculation of other metrics, such as the standard deviation, by filtering the set of days used as input. Another approach uses weights when calculating deviations, giving more weight to the same day of the week to account for weekly variations~\cite{gadaleta2021passive}.  However, efficient computation techniques to deal with multiple contextual variables, as well as their possibly combinatorial explosion, are less often considered. 

Understanding the association with context might be critical for designing interventions for health, and for providing more holistic insights, therefore, including context in the modeling and analysis of behavioral and physiological signals is critical.  

\subsection{Limitations of this study}
This study does not adhere to the methodology of a systematic literature review, and consequently, may be subject to a degree of bias in the selection of papers. The myriad of terms and techniques employed in the field posed a significant challenge to conducting a systematic review. Instead, the primary focus was directed towards identifying and establishing commonalities and distinctions among the various models, as well as their respective applications. The study aims to encompass a wide array of literature to furnish a comprehensive overview of the models pertaining to long-term patterns.

\section{Conclusion}
\label{sec:conclusion}

This review has explored computational models for analyzing longitudinal wearable sensor data in healthcare, focusing on routines, rhythms, and stability metrics. These models offer valuable tools for understanding recurring patterns in individual behavior and physiology, with potential applications ranging from early disease detection to personalized intervention design.

Routines, with their focus on sequences of events, offer valuable interpretability, making them well-suited for designing targeted interventions. By explicitly modeling how specific behaviors unfold over time, routines can reveal actionable insights into the relationship between daily activities and health outcomes. Rhythms, on the other hand, provide a powerful means of uncovering disruptions in biological clocks and gaining insights into underlying physiological processes. Analyzing rhythm characteristics such as period, phase, and amplitude can reveal subtle deviations from healthy patterns, offering valuable information for diagnosis and treatment. Metrics, in turn, provide an easily interpretable and calculable way to quantify the regularity and stability of patterns detected by routines and rhythms. This simplification of complex longitudinal data makes them particularly useful for identifying associations with various health outcomes.

However, each approach has its limitations. Routines heavily depend on accurate event definition, which can be challenging due to the arbitrary nature of time windows and the potential for error propagation from event identification to pattern modeling. While rhythms excel at analyzing periodicity, they can lead to misleading conclusions if rhythms from different processes are compared without considering confounding factors. For instance, disentangling the influences of circadian processes and physical activity on heart rate rhythms requires careful consideration. Metrics, although offering ease of interpretation and calculation, often rely on a single signal modality and a specific timeframe, potentially restricting the scope of analysis.

Despite the progress in analyzing longitudinal wearable sensor data, several promising research directions can further advance the field. Standardized approaches for defining events and selecting time windows are crucial for enhancing the comparability and generalizability of research findings. Currently, the lack of standardization hinders the translation of research findings into practical applications and makes it difficult to compare results across different studies.

Models should also incorporate contextual factors such as weekends, holidays, and weather to improve their accuracy and provide more holistic insights into individual behavior. These factors can significantly influence patterns and ignoring them can lead to biased results and a limited understanding of behavioral variations.

Integrating routines, rhythms, and stability metrics into a unified framework could offer a more comprehensive understanding of longitudinal patterns. Combining the strengths of each approach can provide a richer perspective on the interplay between behavioral sequences, biological rhythms, and overall regularity, potentially leading to more effective interventions and personalized health insights.

The analysis of longitudinal wearable sensor data, through the lens of routines, rhythms, and metrics, holds immense potential for transforming healthcare. These models offer a powerful means of understanding the intricate tapestry of human behavior and physiology, enabling the development of personalized interventions, early warning systems, and a deeper understanding of the dynamic interplay between lifestyle and health. As the field continues to mature, addressing the challenges of standardization, contextual integration, and data quality will be crucial for realizing the full potential of these models. 
However, more research is needed on the standardization of data analysis to make these results generalizable

\bibliographystyle{plain}
\bibliography{main}

\end{document}